\documentclass[submission,copyright,creativecommons]{eptcs}
\providecommand{\event}{ACL2 2022} 
\usepackage{breakurl}             
\usepackage{underscore}           

\usepackage{amssymb}
\usepackage{mathtools}

\usepackage{amsthm}

\usepackage{listings}

\usepackage{fancyvrb}

\usepackage{upquote}

\usepackage{booktabs,tabularx, colortbl}
\usepackage{multirow}
\usepackage{xcolor}
\usepackage{float}

\newtheorem{lemma}{Lemma}
\newtheorem{example}{Example}
\newtheorem{definition}{Definition}

\usepackage{pgfplots}

\usepackage{subcaption}

\usepackage{verbatimbox}

\usepackage{booktabs,tabularx, colortbl}

\definecolor{darkgreen}{rgb}{0.01, 0.75, 0.24}

\newlength{\Oldarrayrulewidth}
\newcommand{\Cline}[2]{%
  \noalign{\global\setlength{\Oldarrayrulewidth}{\arrayrulewidth}}%
  \noalign{\global\setlength{\arrayrulewidth}{#1}}\cline{#2}%
  \noalign{\global\setlength{\arrayrulewidth}{\Oldarrayrulewidth}}}

\title{Verified Implementation of an Efficient Term-Rewriting Algorithm for
  Multiplier Verification on ACL2}

\author{Mertcan Temel
\institute{University of Texas at Austin\\ Austin, TX, USA}
\email{mert@utexas.edu}
}

\def\titlerunning{Multiplier Verification on ACL2} 
\def\authorrunning{M. Temel}

\begin{document}
\maketitle

\begin{abstract}

   Automatic and  efficient verification of multiplier  designs, especially
   through a provably correct method, is  a difficult problem.  We show how
   to utilize a  theorem prover, ACL2, to implement  an efficient rewriting
   algorithm  for   multiplier  design   verification.   Through   a  basic
   understanding of the features and data  structures of ACL2, we created a
   verified  program  that  can  automatically  verify  various  multiplier
   designs much faster than the other state-of-the-art tools. Additionally,
   users of our system have the flexibility to change the specification for
   the target  design to verify  variations of multipliers. We  discuss the
   challenges we tackled during the development  of this program as well as
   key implementation details for  efficiency and verifiability.  Those who
   plan to implement an efficient program  on a theorem prover or those who
   wish to implement our multiplier verification methodology on a different
   system may benefit from the discussions in this paper.

\end{abstract}

\section{Introduction}
\label{sec:introduction}

Integer multipliers are ubiquitous  circuit components that are fundamental
in general-purpose, cryptographic, image,  and signal processors.  They can
be  used   for  various   arithmetic  operations  such   as  floating-point
multiplication, integer  multiplication, dot-product, division,  and square
root.  Consequently, the correctness of  multipliers is an important factor
for the reliability of such systems.

Formal  verification of  integer  multiplier designs  is  still an  ongoing
problem.   Some earlier  methods, such  as BDDs  and BMDs,  can efficiently
verify  array multipliers~\cite{bmd-paper,bdd-paper},  which are  regularly
structured   designs.    However,   these   methods,   as   well   as   SAT
Solvers~\cite{amr-ahmed2016}, do  not scale  for automatic  verification of
more common  multiplier architectures, i.e., Wallace-tree  like multipliers
and Booth Encoding. Their irregular and more advanced structure complicates
the  process for  automatic  tools; therefore,  verification of  industrial
designs             is            carried             out            mostly
manually~\cite{hardin-centaur,jason-baumgartner,conf-date-KaivolaN02,russinoff2019formal,memocode11}. Recent
studies have focused on computer algebra  based methods and they have shown
significant
improvements~\cite{Ciesielski2019UnderstandingAR,dkaufmann-FMCAD19,Mahzoon2018PolyCleanerCY,Mahzoondac2019,Yu2018FastAR}. Some
of  these  tools~\cite{dkaufmann-FMCAD19,Mahzoondac2019} can  verify  large
isolated integer  multipliers in a  shorter amount of time  than previously
reported  tools (e.g.,  256x256-bit multipliers  can be  verified in  a few
hours, and in some cases, minutes).   However, very little effort was given
to verifying the tools  themselves.  Only one tool~\cite{dkaufmann-FMCAD19}
can  generate certificates  to check  the verification  result by  external
proof checkers.   Additionally, specification for designs  is hard-coded in
these tools,  and it  may not  be possible to  verify modified  or embedded
multipliers with these tools~\cite{fmcad21-paper}.

We have  created a  new term-rewriting algorithm  that can  efficiently and
automatically  verify  large  arithmetic   circuit  designs  with  embedded
multipliers~\cite{fmcad21-paper,cav-paper}.   We   have  shown   that  this
algorithm can verify designs with millions  of gates in just a few minutes.
For example,  an isolated  1024x1024 Booth Encoded  Wallace-tree multiplier
can be  verified in 5  minutes.  Other multiplier-centric designs,  such as
dot-product, and various architectures, such as array multipliers, can also
be  verified quickly.   Our  algorithm  scales much  better  than the  best
available  tool~\cite{dkaufmann-FMCAD19}:   it  can   verify  1024x1024-bit
multipliers around 40 times faster on average.

Not only did we create a  more efficient algorithm, but we also implemented
and verified it  using the interactive theorem prover,  ACL2.  This ensures
the user of the  result when a design is claimed  to be correct.  Moreover,
the interactive  system provides  more flexibility to  the user;  e.g., the
ability to decide on the specification  for a design. In our previous work,
we  have shown  that  we  can verify  custom  designs  with control  logic,
multiply-accumulate,   and    dot-product   with   the   same    level   of
automation~\cite{fmcad21-paper}.




Managing a  term-rewriting algorithm of this  scale is not a  trivial task.
Implementing  and verifying  programs using  an interactive  theorem prover
brings  about  unique  challenges  that   one  might  not  encounter  while
developing unverified programs in other  high-level languages, such as Java
and  C++.   Data   structures  in  ACL2  and  its   rewriter  have  certain
constraints,  and  the proof  obligation  bring  about further  challenges.
While programming, we had to consider  that terms are always represented as
trees and are stored in memory as unmodifiable linked-lists.  Additionally,
terms might change so drastically during rewriting that it might become too
difficult  to apply  some rewrite  rules.  In  our previous  work, we  have
described         only         the         term-rewriting         algorithm
itself~\cite{fmcad21-paper,cav-paper}, and some  features of the supporting
rewriter  technology~\cite{rp-paper}.   The  goal   of  this  paper  is  to
elaborate  on the  multiplier-specific implementation  details and  notable
challenges.

This paper is structured  as follows. Sec.~\ref{sec:Preliminaries} provides
the necessary  background information  to follow  the paper,  including the
term-rewriting  methods in  ACL2  and our  algorithm  to verify  multiplier
designs.      Sec.~\ref{sec:implementation-overview}      describes     the
implementation  framework.   Sec.~\ref{sec:challenges}   details  the  main
challenges  we   encountered  during  implementation  and   our  solutions.
{\footnote {Various  demo files showing how  to input a Verilog  design and
  verify it with  our system, and the source code  of the program discussed
  in this paper can be found in the ACL2 community books~\cite{acl2-github}
  under the books/projects/rp-rewriter/lib/mult3 directory.}}

Some portions  of this work  have previously  appeared in the  author's PhD
thesis~\cite{TemelDissertation}. 

\section{Preliminaries}
\label{sec:Preliminaries}

In this  section, we  describe the  basic and relevant  methods to  apply a
term-rewriting algorithm  in ACL2 as well  as our algorithm to  rewrite and
simplify multiplier designs.

\subsection{Methods of Rewrite}
\label{sec:methods-of-rewrite}

ACL2  is  a  programming  language  and an  automated  theorem  prover  for
first-order  logic~\cite{DBLP:books/sp/10/KaufmannM10}.   Users  can  model
systems  and  reason  about  them   using  either  its  built-in  features,
user-contributed   libraries,   or   external   tools   such   as   a   SAT
solver~\cite{Hunt2017IndustrialHA,acl2-2,acl2-1,sol-fgl}.       In      our
multiplier verification project, we are  only interested in basic rewriting
capabilities. A given conjecture to the  system can be rewritten by rewrite
rules,         meta        rules~\cite{hunt-meta},         or        clause
processors~\cite{clause-processors}.

\paragraph*{Rewrite Rules}

Previously proven lemmas  can be stored as rewrite rules  and later be used
to rewrite  terms to  help prove conjectures.   If a lemma  is of  the form
$hyp \implies lhs = rhs$ and it is enabled in the ACL2 system as a rewrite rule,
then the  $lhs$ will  be compared  to terms being  rewritten. If  the $lhs$
pattern can  be unified with  a (sub)term and  $hyp$ is relieved,  then the
term  will  be  replaced  with  the $rhs$  pattern  with  appropriate  term
bindings.


Table~\ref{rw-rule-example} shows an example of how a rewrite rule alters a
term.  The  {\tt defthm} utility is  used to submit conjectures  to ACL2 to
attempt proofs.  We can prove the associativity of summation (see the first
column) using the existing libraries and built-in axioms in ACL2.  When the
{\tt defthm} event succeeds, it automatically  saves the lemma as a rewrite
rule with the given name.  When this  rewrite rule is in the system, we can
apply it  to terms  whenever the  pattern from the  left-hand side  finds a
match. Assume that this is the only enabled rule and we would like to prove
another  conjecture which  contains a  term  given in  the ``Target  Term''
column.  The  rewriter performs  inside-out rewriting.  Therefore,  it will
first dive  into the innermost  term to  search for matching  patterns. The
first match occurs for the following bindings: {\tt a} to {\tt x3}, {\tt b}
to {\tt x4}, and {\tt c} to {\tt x5}. With these term bindings, the term is
replaced using  the right-hand side of  the rewrite rule and  we obtain the
term in  the third  column.  The rule  can find another  match on  this new
term.  After rewriting  this term in a similar fashion,  we obtain the term
in the last column.

\begin{table}[]
  \caption{An example term being modified by an example rewrite rule }
  \label{rw-rule-example}
  \begin{tabular}[t]{| p{0.235\textwidth} | p{0.191\textwidth} | p{0.216\textwidth} | p{0.228\textwidth} |} %
    \noalign{\hrule height 1pt}

    Rewrite Rule & Target Term & After RW1 & After RW2  \\
   \noalign{\hrule height 0.75pt}
    \vspace{-10px}
\begin{Verbatim}[commandchars=\\\{\}]
(defthm sum-assoc
  (equal
   (+ (+ \textcolor{red}{a} \textcolor{red}{b}) \textcolor{red}{c})
   (+ a (+ b c))))
 \end{Verbatim}
 &
\vspace{-10px}
\begin{Verbatim}[commandchars=\\\{\}]
(+ (+ x1 x2)
   (+ (+ \textcolor{red}{x3} \textcolor{red}{x4})
      \textcolor{red}{x5}))
\end{Verbatim}

& \vspace{-10px}
\begin{Verbatim}[commandchars=\\\{\}]
(+ (+ \textcolor{red}{x1} \textcolor{red}{x2})
   \textcolor{red}{(+ x3}
      \textcolor{red}{(+ x4 x5))})
\end{Verbatim}

 & \vspace{-10px}
\begin{Verbatim}
(+ x1
   (+ x2
      (+ x3
         (+ x4
            x5))))
\end{Verbatim}

    \\

\noalign{\hrule height 1pt}

  \end{tabular}
\end{table}

\paragraph*{Meta Rules}

Instead of applying  a fixed pattern from a rewrite  rule, users may define
custom functions  to match patterns  and modify  terms.  We refer  to these
functions  as {\em  meta functions},  and their  associated  rules as  {\em
  meta rules}~\cite{hunt-meta}.    Meta rules   have   associated   trigger
functions. Whenever  the rewriter  encounters an instance  of one  of those
trigger functions,  the associated meta function  is called to  rewrite the
current term. Meta rules can often be  used to achieve better efficiency in
terms of both  applicability and resources (i.e., CPU-time  and memory). In
order to  define meta rules, users  should verify the correctness  of their
meta functions.

There  might  be cases  where  a  system with  only  rewrite  rules is  not
sufficient  to  apply a  term-rewriting  algorithm.   Consider the  example
rewrite  rule given  in Table~\ref{bad-rw-rule-example}.   This rule  shows
that for  the function  {\tt s},  we can remove  duplicate elements  in the
summation argument.  For  this rewrite rule to apply,  candidate terms have
to  match the  exact pattern  from this  rewrite rule,  requiring duplicate
elements to appear at the beginning of the summation for all to be removed.
As seen in the  ``Target Term'' column, this might not  always be the case;
repeated terms {\tt x2} and {\tt x4}  would not be removed by applying this
rewrite rule. We could possibly define  other rewrite rules to update these
terms;  however, there  could be  an  indeterminate number  of elements  in
summations and it is not feasible  to define rewrite rules that would match
all possible patterns.   We can overcome this issue by  defining a function
that can go through the whole  summation term, remove duplicates and return
the desired term.


\begin{table}[]
  \caption{An example that shows a rewrite rule might not be sufficient to apply a term-rewriting strategy }
  \label{bad-rw-rule-example}
  \begin{tabular}[t]{| p{0.44\textwidth} | p{0.49\textwidth}  |} %
    \noalign{\hrule height 1pt}

    Rewrite Rule & Target Term    \\
   \noalign{\hrule height 0.75pt}
    \vspace{-10px}
\begin{Verbatim}[commandchars=\\\{\}]
(defthm s-of-repeated
  (equal (s (i+ a (i+ a b)))
         (s b)))
\end{Verbatim}
    {\em where}

    {\tt (s x) = (mod (ifix x) 2)}

    {\tt (i+ x y) = (+ (ifix x) (ifix y))}

    {\tt (ifix x) = (if (integerp x) x 0)}

 &
\vspace{-10px}
\begin{Verbatim}[commandchars=\\\{\}]
(s (i+ x1
       (i+ {\textcolor{red}{x2}}
           (i+ {\textcolor{red}{x2}}
               (i+ x3
                   (i+ {\textcolor{red}{x4}}
                       (i+ {\textcolor{red}{x4}}
                           x5)))))))
\end{Verbatim}
    \\

\noalign{\hrule height 1pt}

  \end{tabular}
\end{table}

A  simple program  consisting of  two meta functions  that can  remove such
duplicate elements  is given  in Example~\ref{meta-functions}.   The second
function {\tt s-of-repeated-meta-fn} is to be  called every time a new {\tt
  s} instance is created. Then, the  argument of such an instance is passed
to  the  first  function,  {\tt  rm-dp-for-s}.  This  function  recursively
examines all the elements in a  term representing summation.  If there is a
duplicate,  the  function   removes  it;  and  for  all   other  terms,  it
reconstructs  the  summation.   Finally,  the  updated  summation  term  is
returned  to the  second function  and  a functionally  equivalent {\tt  s}
instance  is  created.   This   program  assumes  that  terms  representing
summations   are   lexicographically   sorted  (using   commutativity   and
associativity of summation), ensuring that repeated elements appear next to
each other.

\begin{example}
  \label{meta-functions} 
  Meta functions that can remove duplicates from
  the  summation   argument  of  an  {\tt   s}  instance.
\tt
\begin{Verbatim}[commandchars=\\\{\}]
(define rm-dp-for-s (term)
  (case-match term
    (('i+ x ('i+ y z))
     (if (equal x y)
         (rm-dp-for-s z)
       `(i+ ,x ,(rm-dp-for-s `(i+ ,y ,z)))))
    (('i+ x y)
     (if (equal x y)
         ''0
       term))
    (& term)))

(define s-of-repeated-meta-fn (term)
  (case-match term
    (('s sum)
     `(s ,(rm-dp-for-s sum)))
    (& term)))
\end{Verbatim}
\end{example}

\vspace{10pt}

Verification of meta functions is  often more difficult than rewrite rules.
Not only  do we need to  prove properties about the  interpreted functions,
such as {\tt s} and {\tt i+}, we  also need to show that the meta functions
return equivalent terms. Example~\ref{verify-meta-functions} shows the list
of events to verify the  functions in Example~\ref{meta-functions}. We omit
the  lemmas about  {\tt  s}  and {\tt  i+}  (e.g.,  {\tt s-of-repeated}  in
Table~\ref{bad-rw-rule-example})  for brevity.   To verify  meta functions,
ACL2  requires  users to  create  an  {\em  evaluator} that  can  recognize
functions  in rewritten  terms.   Then,  we prove  a  lemma  for each  meta
function with this  evaluator.  The form {\tt (my-eval  term a)} represents
evaluation  of {\it  {\tt  term}}  for any  binding  alist  {\tt a},  which
represents bindings for any free variable  that might appear in {\tt term}.
The first lemma  {\tt rm-dp-for-s-is-correct} states that  the updated term
by our  function {\tt rm-dp-for-s} will  evaluate to the same  value as the
original term when both appear as an  argument of {\tt s}. The second lemma
states  that the  evaluation of  {\tt term}  will be  the same  when it  is
updated by our second function  {\tt s-of-repeated-meta-fn}.  This lemma is
saved as a meta rule in ACL2  so that this function is triggered whenever a
new instance of {\tt s} is encountered during rewriting.

\newpage
\begin{example}
  \label{verify-meta-functions} 
  Verification of meta functions and the special {\tt defthm} call
    to       register       the        meta       function       as       a
    meta rule
 \tt
\begin{Verbatim}[commandchars=\\\{\}]   
(defevaluator my-eval my-eval-lst ((s a) (i+ a b)))

(defthm rm-dp-for-s-is-correct
  (equal (s (my-eval (rm-dp-for-s term) a))
         (s (my-eval term a))))

(defthm s-of-repeated-meta
  (equal (my-eval term a)
         (my-eval (s-of-repeated-meta-fn term) a))
  :rule-classes ((:meta :trigger-fns (s))))
\end{Verbatim}
\end{example}


\paragraph*{Clause Processors}

Similar  to  meta rules,   clause  processors~\cite{clause-processors}  can
rewrite terms with user-defined functions.   There are two main differences
between meta rules and  clause processors. First, clause  processors do not
require  a   trigger  function,   but  they  rewrite   disjunctive  clauses
representing  a  given conjecture.  Second,  clause  processors may  return
stronger/more general  conjectures than the  input.  For example,  a clause
processor may drop some or all of the hypotheses of a given conjecture.

ACL2  has a  very capable  built-in rewriter  but it  is not  developed and
optimized for  conjectures that can grow  into very large terms.   For this
reason, we developed a custom rewriter, called RP-Rewriter~\cite{rp-paper},
to more efficiently  deal with large terms from  conjectures for multiplier
designs   (e.g.,   Listing~\ref{multiplier-is-correct}).   RP-Rewriter   is
designed and  used as a clause  processor.  This rewriter imitates  some of
the  features  of  ACL2's  built-in   rewriter  but  also  implements  some
optimizations  for large  terms (e.g.,  fast-alist support~\cite{rp-paper})
and some rewriting features that are needed by our term-rewriting algorithm
(e.g.,     {\em     side-conditions}     feature    as     discussed     in
Sec.~\ref{sec:side-conditions}).    RP-Rewriter   is  a   verified   clause
processor,  that  is,  all  the  rewriting performed  on  a  conjecture  is
guaranteed to  be sound.  We  let {\em only} RP-Rewriter  manipulate target
conjectures  for multiplier  designs  and never  the  built-in rewriter  by
default.   An existing  set of  rewrite/meta rules  in ACL2  can be  easily
adapted  to  work  with  RP-Rewriter.  Interested  readers  may  find  more
information      about      RP-Rewriter      in      author's      previous
work~\cite{rp-paper}. That  work details  RP-Rewriter but does  not deliver
in-depth discussions about how  our multiplication framework is implemented
using this rewriter.

Clause  processors can  be used  with other  clause processors  as well  by
concatenating    clause-processor    hints    (see   the    related    ACL2
documentation~\cite{acl2-doc}).   For example,  in cases  where RP-Rewriter
cannot   conclude   a    correctness   proof,   we   can    use   the   FGL
system~\cite{sol-fgl} as a clause processor  to send the simplified term to
an external SAT solver. This  can help generate counterexamples or possibly
conclude the proofs if the given conjecture is correct.


\subsection{The Term-Rewriting Algorithm for Multiplier Designs}
\label{sec:the-algorithm}

The  target designs  for  our verification  algorithm  are various  integer
multiplier  designs  coded  in  a hardware  description  language  such  as
Verilog.  Integer multiplier design algorithms, such as Wallace-tree, build
circuits using  unit adders, such  as half/full-adders, and  vector adders,
such as  a carry-lookahead  adder.  Verilog  designs often  implement these
adders with  a design hierarchy,  that is, the  instances of all  the adder
modules are distinguishable from the rest of the circuit.  Our verification
algorithm takes  advantage of  this property, and  we simplify  and rewrite
adder modules before attempting to verify the target multiplier module.


\begin{definition}
  \label{s-c-def}
  Functions $s$ and $c$ are defined as follows.
  \begin{align*}
    \forall x \in \mathbb{Z} \ s(x) &= mod_2 (x) \\
    \forall x \in \mathbb{Z} \ c(x) &= \Bigl\lfloor {\frac{x}{2}} \Bigr\rfloor
  \end{align*}
\end{definition}

The first  step of our verification  algorithm is to rewrite  the output of
adder  modules  in  terms  of  the  {\tt s}  and  {\tt  c}  functions  (see
Def.~\ref{s-c-def}). For  each distinct adder  module, we prove  that their
output can be represented in terms of {\tt s}, {\tt c} and {\tt +} as given
in  Table~\ref{adder-and-mult-form}.   Since  all  the  adder  modules  are
hierarchical in  the target designs, they  are all replaced by  these forms
when later symbolically simulating the multiplier module.





\begin{table}[]
  \caption{Targeted final forms of the verification algorithm for some modules/functions}
  \label{adder-and-mult-form}
  \vspace{-10px}
  \begin{tabular}[t]{| p{0.21\linewidth} | p{0.22\linewidth} | p{0.22\linewidth} | p{0.22\linewidth} |} 
    \noalign{\hrule height 1pt}
    Function &  $out_2$ & $out_1$ / $c_{out}$ & $out_0$ / $s_{out}$ \\
    \noalign{\hrule height 1pt}
    Half-adder & - & $c(a + b)$ & $s(a + b)$ \\
    \noalign{\hrule height 1pt}
    Full-adder & - & $c(a + b + c_{in})$ & $s(a + b + c_{in})$ \\
    \noalign{\hrule height 1pt}
    Bit-vector

    addition

    $a+b$
          &
            \begin{tabular}[t]{@{}l@{}}
              $s(a_2 + b_2$\\
              $\ +c(a_1 + b_1$ \\
              $\ \ \ +c(a_0 + b_0)))$
            \end{tabular}
          &
            \begin{tabular}[t]{@{}l@{}}
              $s(a_1 + b_1  $ \\
              $\ +c(a_0 + b_0 ) )$
            \end{tabular}
          &
            $s(a_0 + b_0 )$\\
    \noalign{\hrule height 1pt}
    Bit-vector

    multiplication

    $a*b$

&
     \begin{tabular}[t]{@{}l@{}}
                $s(a_0b_2 + a_1b_1 + a_2b_0$\\
                $\ \ \ \ +c(a_1b_0 + a_0b_1$ \\
                $\ \ \ \ \ \ \ \ \ \ +c(a_0b_0)))$
              \end{tabular}
            &
              \begin{tabular}[t]{@{}l@{}}
                $s(a_1b_0 + a_0b_1 $ \\
                $\ \ \ \ +c(a_0b_0 ) )$
              \end{tabular}
            &
              $s(a_0b_0 )$\\
    \noalign{\hrule height 1pt}
  \end{tabular}
  \vspace{-5px}
\end{table}

After  finding  appropriate representations  for  adder  modules, we  start
simplifying the  main multiplier design.  Multipliers  generally consist of
two parts: partial product generation  and summation. We have two different
set of lemmas to rewrite and simplify these segments.

Partial product generation algorithms, such  as Booth encoding, may involve
some optimizations to reduce gate-delay.  These optimizations may result in
very  complex Boolean  expressions. We  rewrite such  terms with  a set  of
lemmas of the  form $lhs$ = $rhs$, where terms  matching $lhs$ are replaced
with        $rhs$       with        appropriate       term        bindings.
Lemmas~\ref{pp-lemma1a},~\ref{pp-lemma1b},~\ref{pp-lemma1c}         perform
algebraic rewriting to reduce the logical expressions into terms involving only $+$,
$-$ and $\land$ operators.

\begin{lemma} \label{pp-lemma1a}
    $ \forall x \in \{0,1\}\ \neg x = 1 - x $
\end{lemma}
\begin{lemma} \label{pp-lemma1b}
    $\forall x, y \in \{0,1\}\ x \lor y = x + y - x y\ $
\end{lemma}
\begin{lemma} \label{pp-lemma1c}
    $\forall x, y \in \{0,1\}\ x \oplus y = x + y - x y\ - x y\ $
\end{lemma}

The majority  or all  of partial  products are passed  to adder  modules as
inputs in multiplier  designs.  Since adder modules are  rewritten in terms
of {\tt  s} and {\tt  c}, we can see  expressions from partial  products as
arguments  of the  {\tt  s} and  {\tt c}  functions.   We simplify  certain
occurrences           of           these          functions           using
Lemmas~\ref{pp-lemma2a},~\ref{pp-lemma2b},~\ref{pp-lemma3a},~\ref{pp-lemma3b}.

\begin{lemma} \label{pp-lemma2a}
  $\forall x, y \in \mathbb{Z}\ s((-x) + y) = s(x + y)         $
\end{lemma}
\begin{lemma} \label{pp-lemma2b}
  $\forall x, y \in \mathbb{Z}\ c((-x) + y) = (-x) + c(x + y)  $
\end{lemma}
\begin{lemma} \label{pp-lemma3a}
  $ \forall x, y \in \mathbb{Z}\ s(x + x + y) = s(y)        $
\end{lemma}
\begin{lemma} \label{pp-lemma3b}
  $ \forall x, y \in \mathbb{Z}\ c(x + x + y) = x + c(y) $
\end{lemma}

The  other  major components  of  multiplier  designs are  partial  product
summation  trees.   Summation  trees  consist  of  a  large  web  of  adder
modules. As we  rewrite adder modules in  terms of the {\tt s}  and {\tt c}
functions, we  derive complex expressions  in terms of these  functions. We
simplify    such    expressions   using    Lemmas~\ref{s-of-s-lemma}    and
~\ref{c-of-s-lemma}.

\begin{lemma}
  \label{s-of-s-lemma}
    $ \forall x, y \in \mathbb{Z}\ s(s(x) + y) = s(x + y)\ $
\end{lemma}

\begin{lemma}
  \label{c-of-s-lemma}
    $ \forall x, y \in \mathbb{Z}\ c(s(x) + y) = c(x + y) - c(x)\ $
\end{lemma}

These lemmas  help reduce expressions  derived from symbolic  simulation of
multiplier    designs   to    a    fixed   final    form    as   seen    in
Table~\ref{adder-and-mult-form}.   We can  convert the  specification of  a
design  into  the same  form  and  conclude  our  proofs with  a  syntactic
comparison.

Additional  details about  this  algorithm  may be  found  in our  previous
work~\cite{fmcad21-paper,cav-paper}, where readers may find a more thorough
explanation  and  examples  as  to   how  this  algorithm  reduces  complex
multiplier designs  to a  fixed form.   These studies  also show  that this
algorithm  is  applicable  to   many  different  design  strategies  (e.g.,
signed/unsigned Booth Encoding; Wallace, Dadda, and other summation trees),
scales almost  linearly with circuit  size and delivers  consistent results
across  different   architectures.   It   can  verify   even  1024x1024-bit
multipliers  or similarly  sized dot-product  designs within  minutes. What
these publications lack, however, are  key implementation details that make
this algorithm efficient and verifiable.

\section{Implementation Overview}
\label{sec:implementation-overview}

An input design needs to be  represented with suitable semantics in ACL2 in
order  to state  a  conjecture  and verify  its  functionality.  Since  our
multiplier verification algorithm utilizes design hierarchy, we use the SVL
system~\cite{svl-books} that  retains hierarchy while simulating  a design,
which makes it possible to rewrite instances of adder modules when they are
used as  submodules.  The  SVL system takes  advantage of  the industrially
used  VL  and  SV  toolkits~\cite{Hunt2017IndustrialHA}  to  parse  Verilog
modules.   This makes  it possible  for the  SVL system  to parse  advanced
modules but it may not support modules with large control logic that can be
found in  industrial designs.  Thankfully,  our setup makes it  possible to
use   other   semantics   as    well~\cite{fmcad21-paper}   such   as   the
industrial-design-friendly SVTV system~\cite{Hunt2017IndustrialHA}.

Before working  on a multiplier  module, we  state rewrite rules  about its
adder modules so that their output can be rewritten in terms of the {\tt s}
and  {\tt   c}  functions.   Listing~\ref{full-adder-is-correct}   shows  a
simplified version of such a rewrite  rule for the SVL semantics. This rule
will  be used  when  we  later symbolically  simulate  the main  multiplier
module.  Instead of {\tt s} and  {\tt c}, we use their logically equivalent
functions {\tt s-spec} and {\tt  c-spec}, which will trigger meta functions
to apply our term-rewriting algorithm.

\begin{lstlisting}[  caption={A simplified  rewrite rule  for a  full-adder
      module.  This rule will be  applied when verifying a multiplier module.},
    basicstyle=\ttfamily,
  label=full-adder-is-correct]
(defthm full_adder_is_correct
  (implies (and (bitp a)
                (bitp b)
                (bitp cin))
           (equal (svl-run (list a b cin) <full_adder>)
                  (list (s-spec (+ a b cin))
                        (c-spec (+ a b cin))))))
\end{lstlisting}

After  creating rewrite  rules  for all  the adder  modules,  we state  the
desired     conjecture     for      the     main     multiplier     design.
Listing~\ref{multiplier-is-correct} shows  an example of such  a conjecture
and the event  to submit to ACL2  in order to have our  system simplify and
prove it.  This conjecture contains two  free variables {\tt a} and {\tt b}
that represent  the two  bit-vectors that  the design  takes as  input. The
left-hand side  is the symbolic simulation  of a design in  SVL format with
these free variables. In the actual  program, {\tt svl-run} returns an {\em
  alist} of  output signals; however,  for brevity, assume that  it returns
the  value of  the only  output signal  here.  The  right-hand side  is the
specification.  We provide  a clause processor hint to  {\tt defthm}, which
will  call   RP-Rewriter~\cite{rp-paper}  to   simplify  and   verify  such
conjectures.     As    discussed   in    Sec.~\ref{sec:methods-of-rewrite},
RP-Rewriter is a  verified clause processor with  certain optimizations for
large terms and it will be used to apply our term-rewriting algorithm.

\begin{figure}[]
\centerline{\includegraphics[scale=0.30]{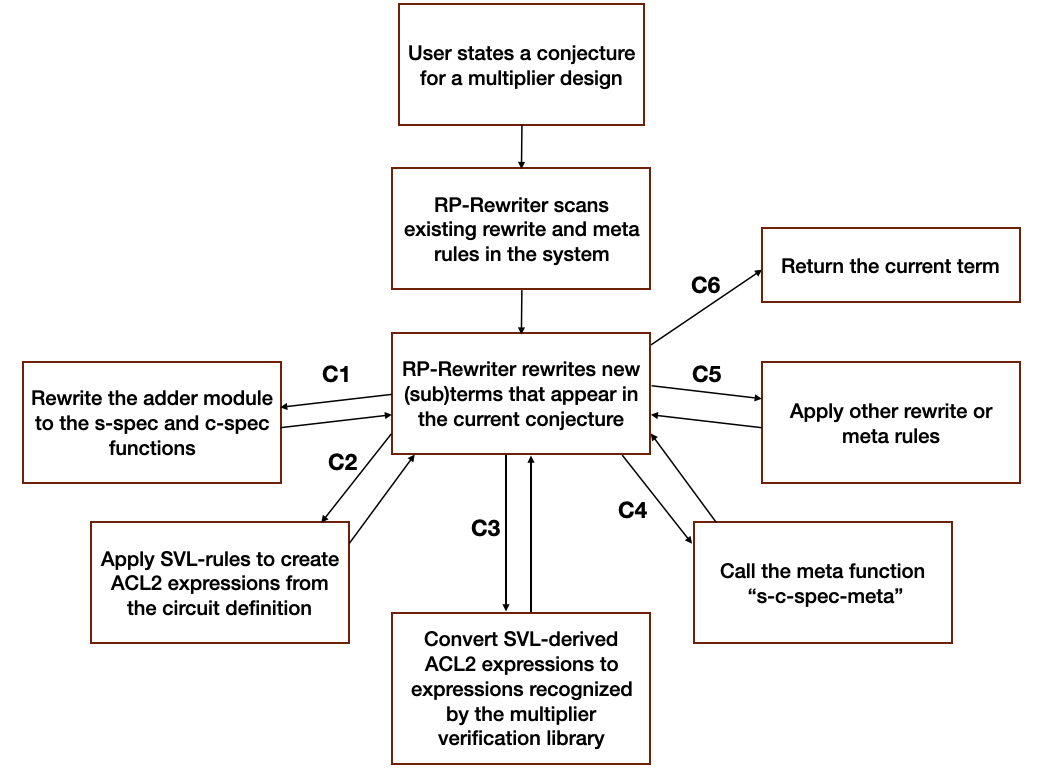}}
\caption{Term rewriting flow when verifying a multiplier design. Transition conditions:
  (C1) the current term is an instance of svl-run-phase of an adder module;
  (C2) An instance of an SVL simulation function;
  (C3) An instance of an ACL2 expression appeared as a result of SVL-rules;
  (C4) An instance of the s-spec or c-spec functions;
  (C5) Some other instance which may be rewritten by some rewrite/meta rules;
  (C6) There is no applicable rule.
}
\label{fig:rw-flow}
\end{figure}

\begin{lstlisting}[%basicstyle=\ttfamily,
  caption={A simplified correctness conjecture for a signed 64x64-bit multiplier with  SVL semantics},
  label=multiplier-is-correct,basicstyle=\ttfamily]
(defthm multiplier_is_correct
  (implies (and (integerp a)
                (integerp b))
           (equal (svl-run (list a b) <signed_64x64_mult>)
                  (loghead 128 (* (logext 64 a)
                                  (logext 64 b)))))
  :hints (("Goal" :clause-processor (rp-cl))))
\end{lstlisting}

Fig.\ref{fig:rw-flow}  shows   the  rewriting   flow  when   simplifying  a
multiplier design conjecture.  Transition  conditions C1-6 are ordered with
respect to their application priority.  Whenever one of these conditions is
met on the  currently rewritten term, the associated  rule(s) apply.  These
conditions and  the rewriting scheme are  developed in a way  that allows a
single pass rewriting  without any expensive global  search/lookup on terms
when  proving multipliers  correct. We  describe each  of these  transition
conditions below, first describing C2 and C3 before C1 in an effort to help
the readers understand our system more easily.


\paragraph*{C2}When the rewriter starts  working on the given conjecture,
such  as  the one  in  Listing~\ref{multiplier-is-correct},  it works  from
inside-out.  It starts with the  {\tt svl-run} instance.  This function and
its subroutines have a logical definition in ACL2.  With these definitions,
we  use  rewrite and  meta  rules  to  quickly  expand such  {\tt  svl-run}
instances into regular ACL2 terms that represent the symbolic simulation of
a given design.  For example, if  a given design implements bitwise logical
NAND of two vectors, {\tt a} and {\tt b}, then these rules would rewrite an
instance of  {\tt (svl-run (list  a b) <dummy-NAND>)} to  {\tt (4vec-bitnot
  (4vec-bitand a b))}.   Here, the {\tt 4vec-bitand}  and {\tt 4vec-bitnot}
functions perform  bitwise operations  on four-valued  ('0', '1',  'X', and
'Z')  bit-vectors.   We will  refer  to  these  rules as  {\em  SVL-rules}.
SVL-rules  are  defined  independently  from  our  multiplier  verification
library and can work with any design  to convert them to regular ACL2 terms
defined with {\em\tt 4vec} functions.

\paragraph*{C3}The resulting terms  from SVL-rules are specific to the  SVL
and its parent libraries. We define a  small set of bridge rules to convert
such terms  into expressions that  our multiplier verification  library can
recognize.   For example,  {\tt  binary-and} is  a  function that  performs
logical AND  on two-valued ('0'  or '1') numbers and  it is defined  in our
multiplier  verification   library.   These   bridge  rules   convert  {\tt
  (4vec-bitand  a b)}  to {\tt  (binary-and a  b)} if  we know  through the
current context that {\tt a} and {\tt b} are two-valued numbers.

\paragraph*{C1}In ACL2,  a more recently defined  rewrite rule has priority
over previous rules (i.e., it will be tried first), which is how we rewrite
instances of  adder modules in terms  of the {\tt s-spec}  and {\tt c-spec}
functions.  When  the SVL functions  simulate a module, it  makes recursive
calls for the  simulation of sub-modules.  For example,  when {\tt (svl-run
  (list  a  b)  <signed\_64x64\_mult>)}  is expanded,  the  rewriter  might
generate many  instances of this form:  {\tt (svl-run (list (f1  a) (f2 b))
  <half\_adder>)}.    When    users   define   a   rewrite    rule   (e.g.,
Listing~\ref{full-adder-is-correct}),   the  rewriter   can  replace   this
instance with {\tt  (list (s-spec (f1 a)  (f2 b)) (c-spec (f1  a) (f2 b)))}
instead of expanding the half adder  module with SVL-rules to generate {\tt
  4vec} functions.   In other words,  such rewrite rules for  adder modules
override  the   transition  condition   C2.   After  such   rewriting,  our
multiplier-specific rules can start applying our term-rewriting algorithm.


\paragraph*{C4}
As new instances of {\tt s-spec}  and {\tt c-spec} are created, a meta rule
implementing        our         term-rewriting        algorithm        (see
Sec.~\ref{sec:the-algorithm}) is  triggered. This  meta rule calls  a large
set of meta functions that  applies the described term-rewriting algorithm.
Our   meta   functions   efficiently   search   for   the   patterns   from
Lemmas~\ref{pp-lemma1a}-\ref{c-of-s-lemma}  and rewrite  terms accordingly.
Some of those  meta functions are given  in Example~\ref{meta-functions}. A
more  detailed discussion  of  these  functions can  be  found in  author's
dissertation~\cite{TemelDissertation}.

When we started  developing this term-rewriting algorithm,  we used rewrite
rules to quickly  test what rewriting strategy works for  our goal and what
does not. After we  came up with the lemmas that we  observed to work best,
we  slowly implemented  the  lemmas  as meta  functions  to achieve  better
proof-time  performance.  Implementing  our  algorithm  as customized  meta
functions has helped us identify bottlenecks in proof-time performance.  We
improved the  time and  memory performance by  optimizing the  functions as
guided by diligent debugging and profiling.

\paragraph*{C5}
When    rewriting   the    left-hand    side   of    the   conjecture    in
Listing~\ref{multiplier-is-correct} finishes,  we end up with  a simplified
term consisting of primarily {\tt s} and {\tt c} instances.  An example for
the    resulting    term     is    given    in    the     last    row    of
Table~\ref{adder-and-mult-form}.   After the  left-hand side,  the rewriter
starts working  on the right-hand  side (specification of the  design).  We
include rewrite rules in our library to rewrite the built-in multiplication
function    (shown    with    {\tt    *})   to    the    same    form    in
Table~\ref{adder-and-mult-form}.   Finally, the  rewriter compares  the two
sides with  a syntactic check.  If  the two sides are  equivalent, then the
conjecture  is proved.   Otherwise,  further  processing (e.g.,  debugging,
utilizing other verification methods on the simplified term) may be needed.

Our system provides flexibility to the user as to how the conjecture can be
defined and  how our multiplier verification  system can be used  for other
designs, e.g., multipliers with saturation.  This means that the conjecture
may have user-defined functions and they may require their own rewrite/meta
rules, or users  may need to improve the system  with additional rules.  To
accommodate for such cases, our  rewriter checks for other applicable rules
in the system, and applies them when they are present.

\paragraph*{C6}
If there  are no  other applicable rules  for the current  term, or  if the
current term is a constant (a  quoted value), then the rewriter returns the
term.





\section{Challenges}
\label{sec:challenges}

Developing a verifiable and efficient software presents certain challenges,
such as limited  flexibility when using various data  structures and subtle
problems resulting from proof obligation.  In this section, we discuss some
of these challenges and our solutions.

\subsection{Data Structure}

\paragraph*{Tree Representation of Terms}

ACL2's built-in  rewriting system,  as well  as RP-Rewriter,  represent and
parse terms  as a tree.  For  example, when parsing and  rewriting the term
{\tt (f  (g x)  (g x)  y)}, the duplicated  subterms {\tt  (g x)}  would be
processed separately  assuming no caching  mechanism is enabled.  This may
not  cause any  concern for  small  terms; however,  a design  that may  be
represented with a directed acyclic  graph (DAG) may yield an exponentially
large term.  Without caution,  this can cause  resource allocation  to grow
exponentially when verifying such designs.

\begin{figure}[]
\centerline{\includegraphics[scale=0.30]{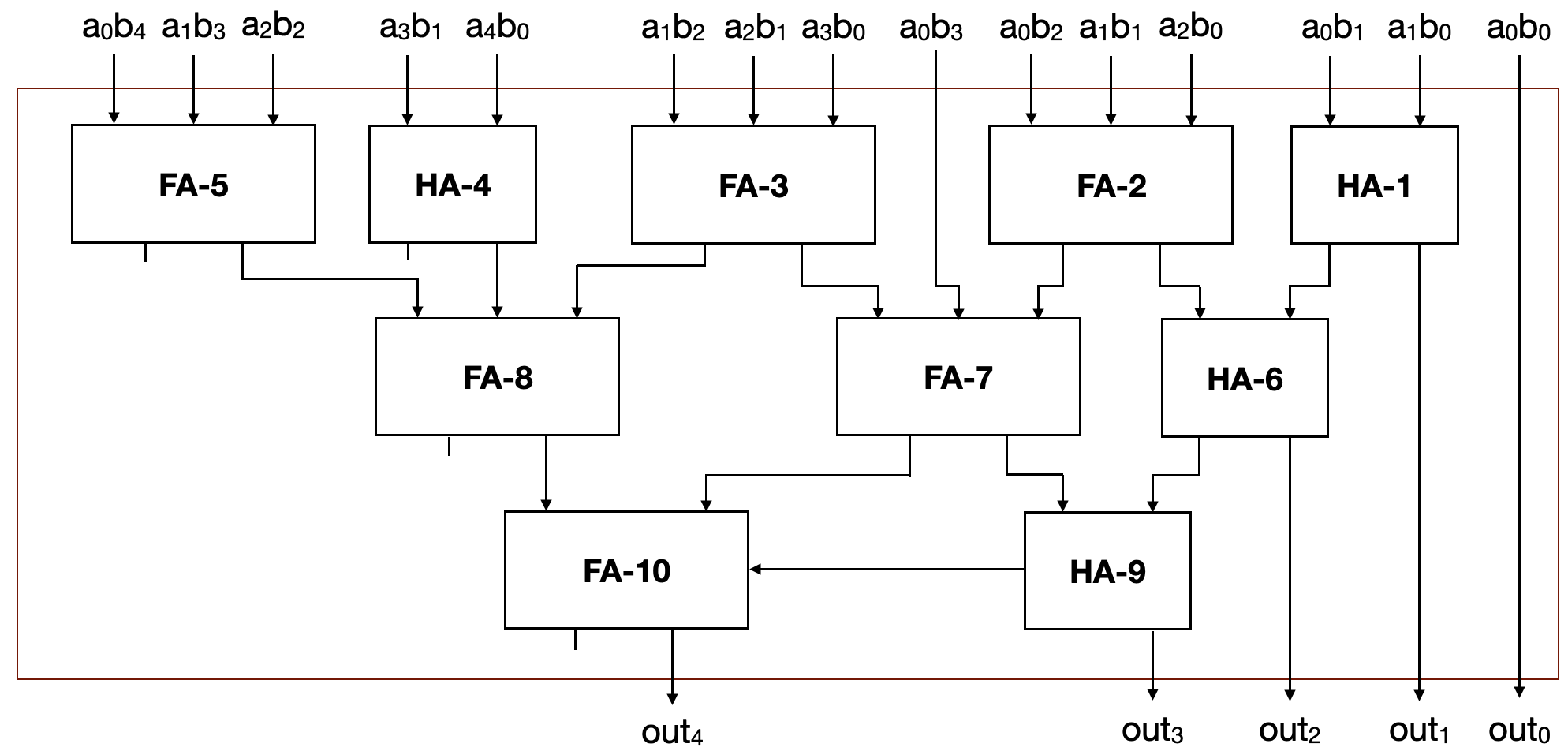}}
\caption{A partial  circuit  diagram  of a  5x5-bit  Wallace-tree  multiplier
  demonstrating its directed acyclic graph structure}
\label{fig:wallace-structure}
\end{figure}

Fig.~\ref{fig:wallace-structure}  shows a  partial  circuit  diagram for  a
Wallace-tree multiplier. We  see that starting from $out_3$,  data from the
inputs  flow  through a  directed  acyclic  graph structure  with  multiple
parents.  When this structure is reduced to the available tree structure to
represent the functionality of each output bit, some subterms are repeated.
This DAG structure persists for  larger multipliers and tree representation
of these designs can grow exponentially with design size.
%

Memoization or a rewrite-cache system may  be used to rewrite flattened DAG
structures to  trees; however, we have  found that it was  not a beneficial
option   for  multiplier   verification  for   numerous  reasons.    First,
efficiently managing a cache table is  a difficult problem.  A large number
of steps are  taken when rewriting multiplier designs, and  caching all the
rewriting may  come at  a substantial  cost for  memory allocation  and run
time.  Developing a smart method to choose  what to cache may or may not be
feasible.  Second,  the memoization system  in ACL2  looks for a  {\em hit}
only  by checking  the  addresses of  objects but  not  their actual  value
because comparing  large terms while  looking for a  cache hit can  be very
expensive.  This can  decrease the number of matches and  the system may be
unable to prevent  repeated rewriting.  ACL2 has a system  to normalize the
addresses of objects,  called {\tt honsing}~\cite{10.1145/1217975.1217992}.
Every time a  new object is created,  this utility registers it  in a table
and  prevents  copies  from  being allocated.   This  can  circumvent  this
memoization  problem but  {\tt honsing}  itself comes  at a  cost as  well:
through our  multiplier verification  experiments, we  found that  it slows
down the program significantly while  increasing memory usage.  This was an
expected outcome as terms change very frequently as rewriting progresses.

Another  workaround  could be  using  a  rewriter  that retains  the  graph
structure (DAG-aware rewriting). However, such a system may bring about its
own challenges, and  memory and time performance may  decline.  The current
tree representation of  terms in our system makes it  easier to develop and
verify meta functions and a rewriter with a more complicated data structure
might make this process more tedious.

In order to  prevent repeated rewriting of the same  terms resulting from a
DAG,  we start  rewriting  designs  with their  original  structure and  we
perform    a    single    pass     of    rewriting    as    discussed    in
Sec.~\ref{sec:implementation-overview}.   The  SVL   format  preserves  the
original DAG structure  of input designs.  We unwind designs  one node at a
time when  rewriting the {\tt  svl-run} instance in our  target conjectures
(see  Listing~\ref{multiplier-is-correct}).  This  system gradually  builds
the final term representing the functionality of each output signal. During
this process, we apply our term-rewriting algorithm when an applicable term
appears.   When the  design has  been  processed completely  using the  SVL
semantics,  we  produce a  term  fully  rewritten  in accordance  with  our
term-rewriting strategy. 

When using other  semantics, the user needs  to be aware of  this issue and
may need to  implement a similar mechanism.  For example,  we implemented a
work-around  for the  SVTV  system~\cite{Hunt2017IndustrialHA} to  overcome
this problem.  The SVTV system represents the functionality of designs with
honsed expressions that are called {\em SVEX}es.  SVEXes are intended to be
interpreted  using  some  memoization  and  honsing  techniques.   For  our
rewriter, these expressions can seem exponentially large because of all the
repeated nodes in  the tree form. We implemented a  rewrite rule mechanism,
that converts  these SVEXes into  a different form of  representation ({\em
  SVEXL},  whose source  code is  located under  books/centaur/svl/svexl in
ACL2  distribution~\cite{acl2-github}).  SVEXL  system  marks the  repeated
nodes in a fashion  that is clearly visible to the  rewriter.  That way, we
obtain a similar mechanism to the SVL when using the SVTV system.

.

\paragraph*{Unmodifiable Linked-lists}

\begin{figure}
  \centering
  \begin{subfigure}{\linewidth}
    \centering
    \includegraphics[scale=0.25]{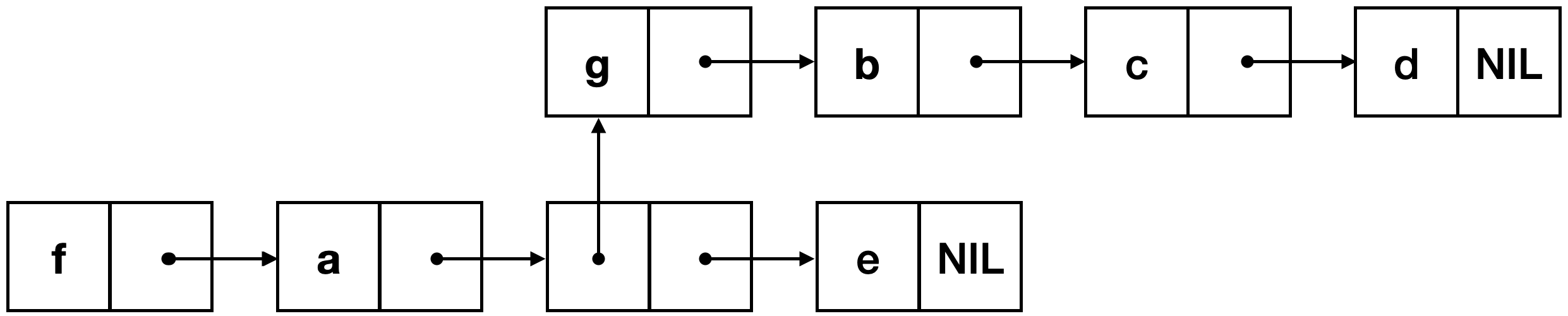}

    \centering
    \caption{Linked-list representation of an example term {\tt (f a (g b c d) e)}}
    \label{fig:linked-list-a}
  \end{subfigure}

  \bigskip

  \begin{subfigure}{\linewidth}
    \centering
    \includegraphics[scale=0.25]{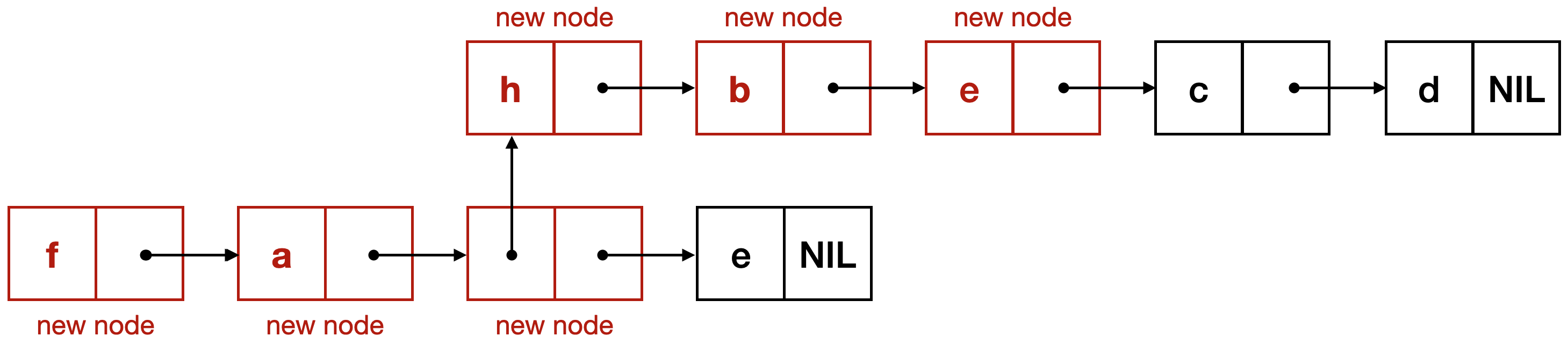}
    \caption{Nodes created after rewriting the example term in (a) to {\tt (f a (h b e c d) e)}}
    \label{fig:linked-list-b}
  \end{subfigure}
  \caption{Representation of ACL2 terms in memory}
\end{figure}

ACL2   terms    are   represented   with   a    pointer-based   linked-list
structure. Fig.~\ref{fig:linked-list-a} shows how an example term {\tt (f a
  (g b c d)  e)} is represented in memory. Each node is  a Common LISP {\tt
  cons} pair, and the first and the second element/pointer in each node can
be accessed with {\tt car} and {\tt cdr}, respectively.

Despite this  linked list structure,  ACL2 forbids many  common linked-list
programming   strategies,  such   as  updating   a  node's   value  without
deallocating/allocating new  nodes. Instead,  whenever a  node in  a linked
list is to  be updated, every node  up to the target node  is discarded and
new nodes are created.   For example, if we would like  to replace the term
{\tt (f a (g b c d) e)} with {\tt (f  a (h b e c d) e)}, then we would have
to  reallocate at  least  6 new  nodes  even though  the  only changes  are
updating   the   value   of   an    existing   one   and   adding   a   new
one. Fig.~\ref{fig:linked-list-b} marks these new nodes.

The reason behind this structure is  basically to keep the ACL2 world sound
while  the  programming  and   verification  procedures  remain  relatively
simple. Assume that  a term is copied  into two places. If  an ACL2 program
updates nodes  in one  of the  copies without  discarding and  creating new
nodes, then  the other copy would  be updated as  well even when it  is not
intended.  It  is not  an easy task  to keep track  of such  situations and
prove  that it  is done  correctly. For  this reason,  ACL2 does  not allow
mutable  linked-lists.  This  programming  limitation can  be difficult  to
adapt for programmers  who make extensive uses of data  structures in other
high-level  languages, such  as C++.   During simple  rewriting, it  is not
trivial to  create O(1)  queues, insert new  elements inside  a linked-list
without  discarding  nodes, and  use  arrays.  Using  {\tt stobj}  or  {\tt
  fast-array} structures to achieve the desired behavior might be an option
but  they  can  complicate  the development  and  verification  process  of
user-defined meta functions.

Profiling of  the rewriter and  meta functions  shows that the  majority of
time  is spent  in functions  that allocate  memory and  only a  negligible
amount of  time is spent  in computation  intensive functions. In  order to
achieve optimal performance, we consider this data structure throughout our
system. For example, we have worked on several optimizations when merging summation
lists. When  rewriting multiplier designs,  our program creates  many terms
that represent summation of partial products as well as {\tt c} and {\tt s}
terms.  Some of the lemmas, such as Lemma~\ref{s-of-s-lemma}, may cause two
summation terms  to be merged.  We  try to keep expressions  in a canonical
form, where we keep all  the summation lists lexicographically sorted.  For
example, if we are  summing the terms {\tt (i+ a (i+ b  c))} and {\tt (i+ d
  e))}, then the final term should be {\tt  (i+ a (i+ b (i+ c (i+ d e))))}.
Simple commutativity rules would use  bubble-sort to create the final form.
At every  step of  this sorting, many  nodes would be  discarded and  a new
linked-list would be created.  We can, however, use a meta function to {\em
  merge} two  lists by iterating  the two lists  only once because  we know
that the input summation lists are always sorted.

Additionally, we divide the summations into three different groups for each
{\tt  s}, {\tt  c}, and  partial product  terms. For  instance, instead  of
defining  the function  {\tt c}  as  {\tt (c  args)} where  {\tt args}  may
represent the summation term of the arguments; we define {\tt c} as {\tt (c
  s-args pp-args  c-args)}, where {\tt  s-args} may represent  summation of
only {\tt s}  arguments, {\tt pp-args} only partial  product arguments, and
{\tt c-args} only {\tt c} arguments (note that in the actual implementation
the {\tt s} and {\tt c} functions  also have another argument {\em hash} as
discussed  in  Sec.~\ref{sec:hash} but  it  is  omitted in  this  section).
Separating elements to such different  lists enables the program to discard
and allocate fewer nodes when  inserting an element to summation arguments.
Such  optimizations  that  focus  on  the  linked-list  structure  and  its
limitations   can  reduce   memory   allocation   substantially.   In   our
experiments, we have seen that  our optimizations for summation lists could
improve  the  performance  (both  memory allocation  and  time)  by  almost
100-fold for some Booth-Encoded 64x64-bit multipliers.

%



\subsection{Side-conditions}
\label{sec:side-conditions}

Throughout our  rewriting, it is  often the case  that terms may  change so
extensively  that it  may become  very difficult  to prove  some properties
about   them.   In   our  previous   work~\cite{rp-paper},  we   introduced
RP-Rewriter's  side-conditions   feature  that  enables  users   to  attach
properties to terms  and retain them throughout rewriting.  This  is one of
the key features of our term-rewriting  system. In this section, we discuss
in detail  how we use  the side-condition feature during  multiplier design
correctness proofs.

The  full/half-adder modules  from Fig.~\ref{fig:wallace-structure}  can be
rewritten in terms of  the {\tt s} and {\tt c}  functions through a rewrite
rule  of the  form given  in Listing~\ref{full-adder-is-correct}  (they are
actually  rewritten  to their  logical  equivalent  {\tt s-spec}  and  {\tt
  c-spec} functions to  trigger meta rules, but we will  refer to only {\tt
  s} and  {\tt c}  functions in  this section).  This  rule has  {\tt bitp}
hypotheses for each input signal, which  indicates that we can rewrite unit
adders in terms  of {\tt s} and  {\tt c} only when the  inputs satisfy {\tt
  bitp}.     Rewriting    starts    from     the    top    of    the    DAG
(Fig.~\ref{fig:wallace-structure})   and  finishes   towards  the   output.
Therefore, we first  apply this rewrite rule for {\tt  FA-2} and {\tt FA-3}
modules.   Then, we  move  to {\tt  FA-7}.  When  attempting  to apply  the
rewrite rule for {\tt FA-7}, we have the following term bindings:
\begin{itemize}
\item[  ] {\tt a} to {\tt (c (+ a0b2 a1b1 a2b0))} (carry output from {\tt FA-2}),
\item[  ] {\tt b} to {\tt a0b3},
\item[  ] {\tt cin} to {\tt (s (+ a1b2 a2b1 a3b0))}  (sum output of {\tt FA-3}).
\end{itemize}
We can  easily relieve the {\tt  bitp} hypotheses of this  rewrite rule for
these bindings, and for the carry output of {\tt FA-7} we obtain:
\begin{verbatim}
  (c (+ (c (+ a0b2 a1b1 a2b0))
        a0b3
        (s (+ a1b2 a2b1 a3b0))))
\end{verbatim}
This term will be automatically  simplified by our term-rewriting algorithm
as  described   in  Sec.~\ref{sec:the-algorithm}.    In  this   case,  only
Lemma~\ref{c-of-s-lemma} will apply. The resulting term is:
\begin{verbatim}
  (+ (c (+ (c (+ a0b2 a1b1 a2b0))
           a0b3
           a1b2 a2b1 a3b0))
     (- (c (+  a1b2 a2b1 a3b0))))
\end{verbatim}
This term will be one of the  inputs when our program tries to rewrite {\tt
  FA-10}.  This  time, it  will not be  as easy to  relieve the  {\tt bitp}
hypothesis.  Even  if we came  up with a smart  rule/set of rules  to prove
that this term satisfies {\tt bitp}, terms can change so significantly that
the cost of backchaining can be very high.

On the other hand, we know that  the resulting forms for each output signal
of unit  adders, such as  the carry  output {\tt (c  (+ a b  cin))}, always
satisfy    {\tt    bitp}.      We    use    RP-Rewriter's    side-condition
feature~\cite{rp-paper} to  remember this  property while we  rewrite adder
modules    in     terms    of     {\tt    s}     and    {\tt     c}    (see
Listing~\ref{full-adder-is-correct}).   To  attach the  side-condition,  we
rewrite the carry output to {\tt (rp 'bitp (c (+ a b cin)))} instead, where
{\tt rp} is logically defined as  an identity function always returning the
second argument.  RP-Rewriter has an invariant  for such {\tt rp} terms and
it {\em  r}etains the {\em p}roperty  that the term in  the second argument
always  satisfies {\tt  bitp} no  matter how  much it  might change  later.
Therefore, when we rewrite the {\tt FA-7} instance, we obtain:
\begin{verbatim}
  (rp 'bitp (c (+ (c (+ a0b2 a1b1 a2b0))
                  a0b3
                  (s (+ a1b2 a2b1 a3b0)))))
\end{verbatim}
for the  carry  output.  Our program  will  rewrite this  term
 the same way, and  but this time we obtain:
\begin{verbatim}
  (rp 'bitp (+ (c (+ (c (+ a0b2 a1b1 a2b0))
                     a0b3
                     a1b2 a2b1 a3b0))
               (- (c (+  a1b2 a2b1 a3b0)))))
\end{verbatim}
This term will  be one of the  inputs to {\tt FA-10}, and  the rewriter can
quickly relieve the {\tt bitp} hypothesis with the attached side-condition.
With  this system,  we can  relieve the  hypotheses for  all adder  modules
without having to backchain or worry about using any extra rules about {\tt
  bitp}.   The   side-conditions  feature  presents  an   O(1)  and  easily
implementable solution.


\subsection{Comparison of Large Terms}
\label{sec:hash}

As  multiplier designs  are  simplified, large  terms  might be  frequently
compared against  each other either  for lexicographical sorting,  to apply
lemmas such  as Lemma~\ref{pp-lemma3a},  or to  cancel terms  in summations
(i.e.,  $a +  (- a)  =  0$).  Comparing  two  large terms  might consume  a
considerable   amount  of   proof-time.    We  have   implemented  a   {\em
  term-hashing} mechanism  to overcome this  issue and quickly  compare two
large terms.

For the {\tt s}, {\tt c} functions, we add a logically extraneous argument,
a hash value calculated with  respect to immediate subterms.  The signature
of these functions are  {\tt (s hash args)} and {\tt  (c hash args)}, where
{\tt hash} is  ignored in the logical definitions. The  other argument {\tt
  args}  consists  of  summation  of  other  {\tt  s},  {\tt  c}  and  {\tt
  binary-and} (from partial products) instances.  One of our meta functions
traverses the  terms in {\tt args}  and using their hash  values (all terms
are expected to  have a hash value),  it calculates a value  for the parent
{\tt s} and  {\tt c} instances. We follow a  similar procedure to calculate
hash values for {\tt binary-and} instances as well.

These hash values help our program  compare two terms much more rapidly. If
the two hash-codes are different, the  program never dives into large terms
and the terms are quickly known to be not syntactically equivalent.  If the
task is  to lexicographically sort  terms, then  it only compares  the hash
values.  If the  hash values are the  same, it may mean that  the two terms
are syntactically equivalent, then the  program dives into the subterms and
compares the whole terms to each other to see if that is indeed the case.

Table~\ref{hash-table} shows  our experiment results when  the term-hashing
feature  is enabled  and  disabled.  We  have tested  the  feature for  110
different  benchmarks~\cite{homma,sca-genmul,tem-genmul} for  various sizes
and   different  architectures:   Wallace-tree   like  multipliers,   array
multipliers,  simple  partial  products  (SP)  and  Booth  Encoded  partial
products (BP). Tests were performed on a iMac Intel(R) Core(TM) i7-4790K CPU
@ 4.00GHz with 32GB system memory.   When disabled, we force hash values to
be 0 for all terms. We see that this feature helps the program scale better
with growing  design size  and it  has resulted in  92\% (13x)  speed-up on
average for 256x256-bit multipliers.

\begin{table}[]
  \caption{Average  proof-time  results  for   a  total  of  110  different
    benchmarks for  various multiplier sizes when  the term-hashing feature
    is enabled and disabled}
 \label{hash-table}

 \begin{tabularx}{0.992\textwidth}{|p{0.2\textwidth} |
     >{\centering\arraybackslash}
     p{0.1\textwidth} |
     >{\centering\arraybackslash}
     p{0.1\textwidth} |
     >{\centering\arraybackslash}
     p{0.1\textwidth} |
     >{\centering\arraybackslash}
     p{0.1\textwidth} |
     >{\centering\arraybackslash}
     p{0.1\textwidth} |
     >{\centering\arraybackslash}
     p{0.1\textwidth} |}

   \noalign{\hrule height 1pt}

  \multirow{2}{*}{Term-hashing} & \multicolumn{2}{c|}{64x64}
  & \multicolumn{2}{c|}{128x128} & \multicolumn{2}{c|}{256x256} \\
  \Cline{1pt}{2-7}
  \noalign{\vspace{1pt}}
  & SP  & BP & SP & BP & SP & BP \\

  \noalign{\hrule height 1pt}
  Disabled (secs) & 1.02 & 2 & 8.12 & 21.15 & 101.9 & 256.6 \\
  \noalign{\hrule height 1pt}
  Enabled (secs) & 0.53 & 1.04 & 2.22 & 4.02 & 11.8 & 18.7 \\
  \noalign{\hrule height 1pt}
  Speed-up & 1.9x & 1.9x & 3.6x & 5x & 8.6x & 13.7x \\
  \noalign{\hrule height 1pt}
\end{tabularx}

\end{table}



\section{Final Words and Conclusions}

We  have  implemented  a  term-rewriting algorithm  for  multiplier  design
verification  on   an  interactive  theorem  prover.    Our  implementation
outperforms other state-of-the-art tools for formal verification of integer
multipliers.   In  our  previous  works~\cite{fmcad21-paper,cav-paper},  we
introduced this  term-rewriting algorithm  and showed  that it  scales very
well with growing circuit sizes (i.e., proof-time grows 4--6 times when the
circuit  grows  ~4  times);  its  performance  is  very  consistent  across
different  benchmarks  as  tested   for  over  20  different  architectures
including  various  Wallace-tree  like and  Booth  Encoded  signed/unsigned
multipliers;  it   is  so   efficient  that  verifying   various  64x64-bit
multipliers  takes   less  than  2   seconds  on  average,   and  verifying
1024x1024-bit multipliers takes around  5 minutes.  Additionally, since our
method  is  applied  using  a  generic rewriter,  it  delivers  a  familiar
interface and  flexibility for  modifications; therefore it  allows designs
with different specifications to be verified, and different semantics to be
used.   This  tool  has  been  used to  verify  various  designs  including
multiply-accumulate,    dot-product,     multipliers    with    saturation,
round-and-scale, truncated or right-shifted multipliers, as well as integer
multipliers  used in  floating-point  designs.  Even  though  we have  only
described its use with the SVL system in this paper, our tool can work with
the  industrially  capable  SVTV  system,  and  has  been  able  to  verify
real-world     designs     at     Intel     Corporation     and     Centaur
Technology~\cite{fmcad21-paper}.

Other            systems             with            the            closest
performance~\cite{dkaufmann-FMCAD19,Mahzoondac2019} do not scale as well as
our program, extending the proof-time  to hours for large multipliers. They
may deliver inconsistent results for  different architectures, and they are
developed for isolated multipliers only.   Even though it might be possible
to include these tools in  other frameworks, embedded multipliers often are
not  constructed   with  isolated   multipliers~\cite{fmcad21-paper}.   For
example,  a multiply-accumulate  design (MAC)  might be  built as  follows.
Partial  products  are created  for  multiplicand  and multipliers.   Then,
instead of  summing the partial  products and obtaining  the multiplication
result  first and  adding  the addend  later, the  addend  and the  partial
products are  summed together  in the  same summation  tree (e.g.,  a Dadda
tree) to calculate the final result of  the MAC.  It may not be possible to
easily extract  an isolated multiplier instance  from such a design  to use
the competing tools.  This integrated  design procedure can be followed for
complex chip  designs to reuse  a large multiplier component  for different
operations,       such       as        dot-product       and       parallel
multiplication~\cite{fmcad21-paper}.   Similarly,  users   may  not  easily
extract an isolated  multiplier from such designs and therefore  may not be
able  to  use  tools  developed  only for  isolated  multipliers  in  their
framework.

Our  implementation in  ACL2 provides  substantially better  results and  a
soundness guarantee. However, developing a competitive and verified program
on  a theorem  prover  has  presented its  own  challenges. Our  experience
teaches  that  understanding the  basic  methods  of term-rewriting  (e.g.,
rewrite rules,  meta rules, and clause processors),  data structures (e.g.,
tree representation of terms, unmodifiable linked-lists), and some rewriter
features (e.g.,  side-conditions) are essential for  creating and verifying
such a system.   Our implementation of a  multiplier verification algorithm
serves as an  example of how an  interactive theorem prover can  be used to
create  large-scale  efficient and  sound  programs  that can  successfully
compete  with   other  state-of-the-art  tools  programmed   in  high-level
languages.

\nocite{*}
\bibliographystyle{eptcs}
\bibliography{acl2-22.paper}
\end{document}